\documentstyle[epsfig]{elsart}

\newcommand{ \Cfour}{C$_4$F$_{10}$ }

\begin{document}
\begin{frontmatter}

\title {The ring imaging  \v{C}erenkov detector
       for the BRAHMS Experiment at RHIC}

\author { R. Debbe}, 
\author {S. Gushue}, 
\author {B. Moskowitz},
\author {J. Olness}, 
\author {F. Videb\ae k}

\address{Brookhaven National Laboratory, 
             Upton, New York 11973, USA}

\begin{abstract}
 A ring-imaging \v{C}erenkov counter, to be  
read out by four 100-channel PMT's, is a key element of the BRAHMS Experiment.
 We report here the most recent 
results obtained with a prototype 
 tested at the BNL AGS using several radiator gases, including
the heavy fluorocarbon \Cfour. 
Ring radii were measured for different particles $(\pi^-, \mu^-, e^-)$
for momenta ranging from 2 to 12 GeV/c employing pure \Cfour as radiator.

\end{abstract}
\end{frontmatter}

\section{The BRAHMS Experiment}

\begin{figure}[h]
\begin{center}
\makebox{\epsfig{height=9.5cm,
         file=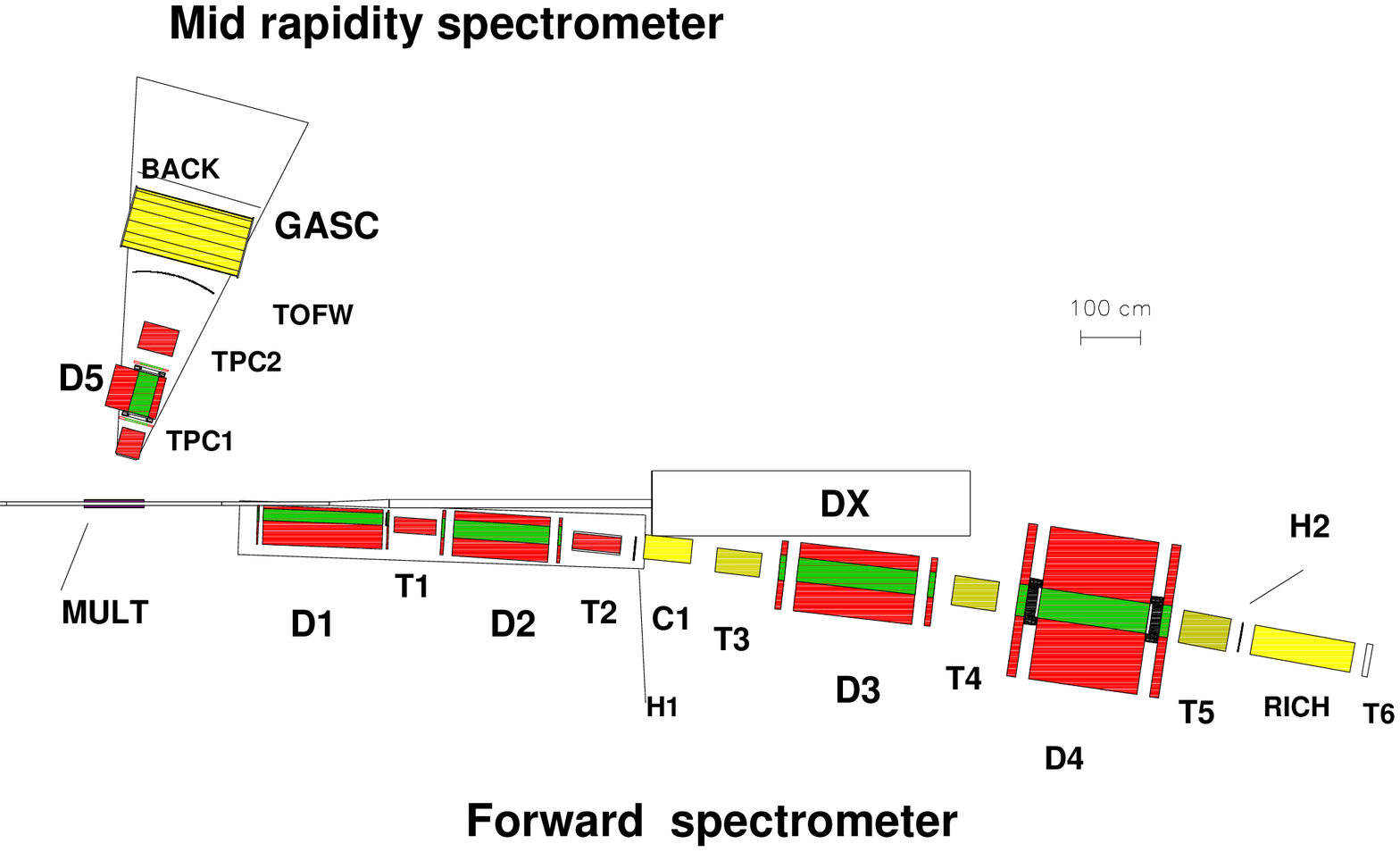}}
\end{center}
\caption{Layout of the BRAHMS spectrometers. Each spectrometer can be rotated
independently about the common vertex. The labels D1 .. D4 indicate the four
dipole magnets of the Forward arm. H1 and H2 are the time-of-flight hodoscopes
placed at 9 and 19 m from the interaction vertex, respectively. T1 and T2 are
TPC's, while T3 ... T6 are wire chambers. The forward arm is shown at $\theta = 2^{\circ}$ and the midrapidity
arm at 75 degrees.}
\label{fig:Brahms_layout}
\end{figure}

	The BRAHMS Experiment at RHIC consists of two magnetic spectrometer
arms that will
survey particle production in Au - Au collisions at 100 GeV/c per nucleon.
The goal of the experiment is to study the system formed by the two 
colliding ions in two regions of rapidity. A schematic description of the
BRAHMS setup is shown in Figure \ref{fig:Brahms_layout}. The Midrapidity Arm
will probe the particles emerging near $ y \approx 0$, corresponding to 
spectrometer
angles between $30^{\circ}$ and $90^{\circ}$  with respect 
to the beam line.
The Forward Arm will study particles of higher rapidities, $y \leq 4$,
corresponding to spectrometer  angles ranging from
$2^{\circ}$ to $30^{\circ}$.

The particle identification is accomplished  in 
each spectrometer
with a combination of Time of Flight arrays 
and \v{C}erenkov counters. Of particular interest for this presentation is the
Particle Identification system (PID) of the Forward Arm spectrometer, 
which must identify charged particles with momenta ranging from
1 to 25 GeV/c. 

Particle identification at high momentum is done with the second Time 
of Flight wall H2 and the Ring Imaging \v{C}erenkov detector RICH.  
Assuming the expected time resolution of $\sigma = 75$ psec, H2  will be 
able to separate 
 pions from kaons up to 5.8 GeV/c.
and kaons from protons up to 8.5 GeV/c. The function of the RICH detector
is to extend the particle identification up to 25 GeV/c. Some considerations
of its design are given in the following subsection.

\section{Design of the RICH detector for BRAHMS}

\begin{figure}[h]
\begin{center}
  \makebox{\epsfig{file=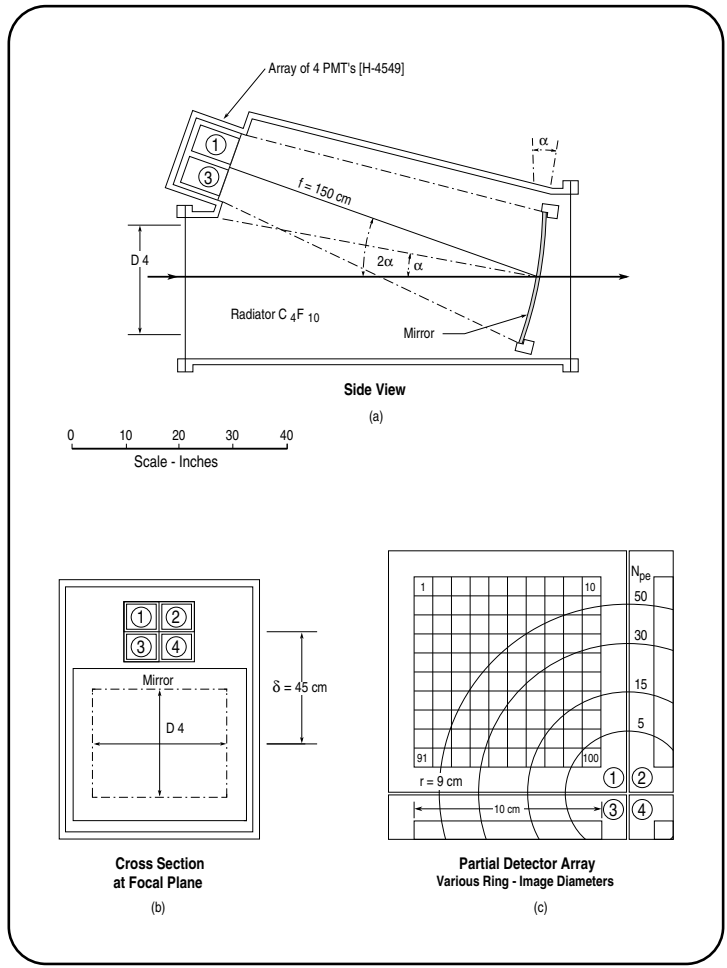,height=12.5cm}}
\end{center}
\caption{Schematic outline of RICH detector}
\label{fig:rich1}
\end{figure}

Figure \ref{fig:rich1} shows three aspects of the proposed detector, 
indicating various components and dimensions. The lateral dimensions of
the radiator volume are somewhat larger than the aperture of D4 
$(h \times w = 40 \times 50 cm^2)$. The radiator has L = 1.5 meters of 
 \Cfour, which at $20^{\circ}$ C and 1.25 atm. pressure has an index of
refraction  n = 1.00185. The spherical mirror has a focal length of 1.5 m
and is rotated by an angle $\alpha = 8^{\circ}$, thus shifting the ring
image (by $2\alpha$) to a focal plane outside of the volume illuminated by
the direct particle flux.

The photon-imaging array consists of four Hamamatsu R-4549-01 detectors,
placed as indicated, defining an image plane some $22 \times 22 cm^2$.

The expected performance of this detector, in terms of particle resolution
and efficiency, have been outlined previously, \cite{Brahms},\cite{memo},
and are supported by the results of the prototype tests presented in the 
following section and also in Ref. \cite{ourNIM}.

Simulations have been performed using event generators appropiate for
Au-Au collisions, in order to determine the multiplicities expected for the 
various detectors of the Forward arm.
Table \ref{table:pid_richrates} shows the pertinent data for the RICH detector,
indicating typically one or at most two charged particles per event.

\begin{table}[h]
 \begin{tabular}{|c|c|c|} 
    \hline
          & Hits per event   & Hits per event  \\
          & $p$=7.5-15 GeV/c & $p$=15-30 GeV/c \\ \hline
Primary   &        0.9       &   0.29          \\
Secondary &        0.5       &   0.39          \\
Above threshold &  0.8       &   0.30          \\ \hline
 \end{tabular}
\caption{Average charged particle hits in the RICH detector}
\label{table:pid_richrates}
\end{table}

\section{The Prototype Detector}
The prototype RICH counter is shown schematically in figure 3 of Ref. \cite{ourNIM}.
The prototype is similar in general design to that shown in figure 
\ref{fig:rich1},
but with the optical system rotated by $90^{\circ}$ about the beam axis, such
that the reflection angle is now in the horizontal plane.

The detector housing is constructed as a rectangular box, with 2 cm thick
aluminum walls, having inner dimensions of 
 127$\times$64$\times$46 cm$^3$ ($l \times w \times h$). The construction is
such that, using gasket and o-ring seals, the pressure of the radiator gas
may be safely varied from 0 to 1.5 atm absolute. The particle beam 
enters and exits through Mylar windows 0.25 mm thick and 15 cm in diameter,
at opposite ends of the long dimension. A 15-cm diameter spherical mirror,
of focal length $f$ = 91.4 cm, is situated at a radiator
distance of $L$ = 114 cm, rotated by  
 $\alpha$ = 8$^{\circ}$. A single 100-channel PMT is centered at 
$\alpha = 16^{\circ}$ at the 91 cm distant focal plane.

The photon detector was a single Hamamatsu R-4549-01, one element of the 
four proposed for BRAHMS. This PMT is a 20-stage device having a $10 \times 10 cm^2$ photocathode, with the segmentation defined by a $10 \times 10 $ array of
$0.9 \times 0.9 cm^2$ anode elements. At the full operating voltage of 2500
Volts, it provides a current gain of $2\times 10^7$ producing single-photon
pulses large enough to be fed directly to an ADC or timing electronics.

This specific module employs a special focussing electrode (between cathode and
first dynode) which results in a rectangular flat-topped response function
$10 \times 10 mm^2$, with a very sharp fall off at the edges. While the 
photocathode itself is quite uniform, the gain falls off a bit (50\%) near 
the edges of the array, and even more at the four corners. However the
signal remains clean, and so the gain can be compensated in software. 

Four identical drift chambers, placed in pairs upstream and dowstream of
the prototype  
 counter, provided tracking for determining the particle 
trajectory.  
Thus, the expected position of the center of
each RICH ring on the phototube is determined to a resolution of 
approximately 500 $\mu$m in both the $x$ and $y$ directions.

\section{Measurements}

In previous tests \cite{ourNIM}, time limitations 
did not permit a complete
filling of the radiator volume with pure \Cfour. Our filling system was 
based on
the difference (factor of six) in the  molecular weights of the fluorocarbon 
and argon, with the heavy gas 
 displacing the lighter one as it is brought into the bottom of 
the detector. In reality we found that even though the flow of \Cfour into the
detector was slow, some mixing occurs and 100\% concentrations cannot be
reached in a single cycle. The measurements reported in our previous 
publication were done with $\approx 70\% $ \Cfour and $30\% $ Argon.  
This time we simply evacuated the radiator volume completely and 
then filled with pure \Cfour. 

\begin{figure}[h]
\begin{center}
\makebox{\epsfig{height=11.5cm,angle=90,file=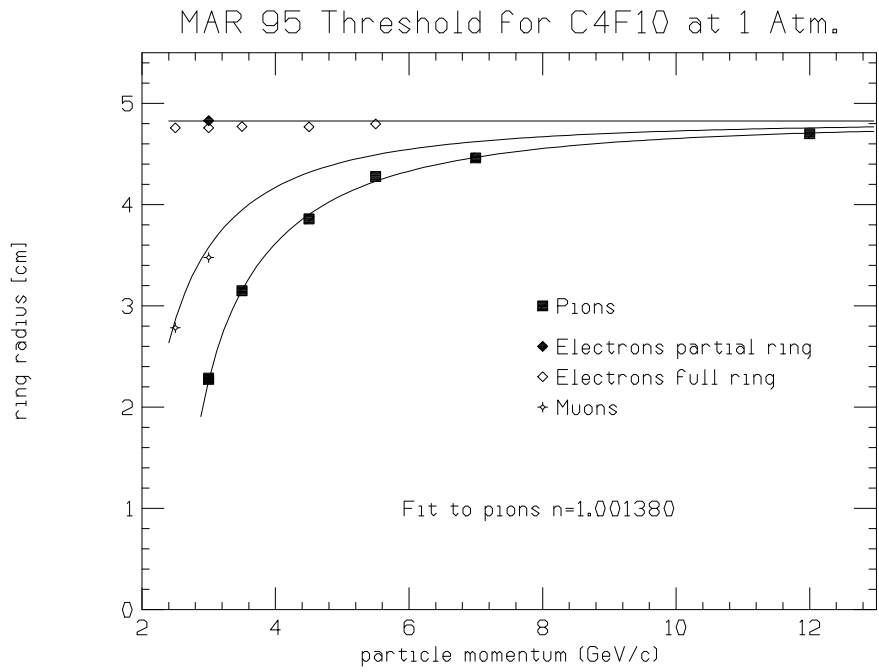}}
\end{center}
\caption{The mean radius from event-by-event fits to rings in the
	 prototype RICH counter filled with a C$_4$F$_{10}$,
	  for different particle species.  The solid curves
	 show the expected radii for an index of refraction of
	 $n=1.001380$} 

\label{fig:momentum_scan}
\end{figure}

Measurements were made for several particle momenta, over the range 
$2 \leq p \leq 12 ~GeV/c$, using the (negative) secondary beam from the
BNL AGS accelerator. (The momentum calibration for the beam-line was
established in a separate measurement, employing time-of-flight techniques, 
to an accuracy of better than 0.5\%). For each setting, the tracking 
information from four
drift chambers was used to project the expected ring centers onto the 
PMT matrix. Given this information, a ring-fitting algorithm was employed
to find and fit the ring radius on an event by event basis. 
The results are shown in 
Fig.~\ref{fig:momentum_scan}, which plots observed ring-radius versus 
beam momentum
for particles thus identified as electrons, muons, and pions.
The solid curves are calculated for the indicated particles, with
the focal length and index of refraction as free parameters. Both results
can be seen to agree well with the known value,$f = 91.4~ cm$, and the
calculated index of refraction (at $20^{\circ}$ C, 1 atm $\approx 400$ nm)
n = 1.001379 (this index is calculated using measurements in the liquid phase
reported in \cite{Ypsi}). 

We have investigated the possibility that for larger rings
(r$\geq$4.5~cm) edge-effects may be of importance, resulting in extracted
radii somewhat too small.  For this purpose, the spherical mirror was
further rotated by an additional 1$^\circ$ to $\alpha$=9$^\circ$.
The projected ring center was therefore moved by 3.4~cm, such that one
arc of the ring image was well inside the PMT's photocathode area.  The
results for this comparison are shown in Fig. \ref{fig:momentum_scan} for 
the electron data at
3~GeV/c.  As can be seen, the effect produces, at most, a minimal
displacement in the direction that might have been expected.

As remarked previously \cite{ourNIM}, the direct determination of the number 
of photons detected for a given event is precluded by the exponential shape
of the PMT single-photon response function.  Instead, we have measured
(over a large number of events) the statistical variation in the total
charge collected per event.  The resultant distribution
($\approx$Gaussian) is then fitted to determine the measured mean
($\mu$)
and standard deviation ($\sigma$).  The number of detected photons (n)
may then be estimated as $n=(\mu/\sigma ')^2$, where $\sigma '$ is
deduced by unfolding the exponential detector response function.  We
thus estimated (for 12 GeV/c pions) that $n=26 \pm 4$, corresponding to $N_0 =
89\pm14~cm^{-1}$.

\section{Summary}

The predicted behavior of the RICH detector proposed for BRAHMS has been
investigated and confirmed by these additional measurements with the
prototype detector, which were carried out under carefully controlled
conditions.  The imaging capabilities (see Fig.~3) and figure of merit
$(N_0)$ are found to be in good agreement with expectations based on the
design and performance of the individual components.

% Acknowledgements
\ack

This work was supported by
U.S. Department of Energy contract number DE-AC02-76CH00016,
in part through the R.\&D. funds of the RHIC Project,
and we thank T.~Ludlam for his encouragement in this enterprise.

%FIGURES
\newpage
%Figure 1 BRAHMS layout

%figure 2 BRAHMS RICH

% Figure 3		momentum_scan

\end{document}